\documentclass[preprint,authoryear]{elsarticle}
\usepackage{graphicx,epsfig,float,rotating,setspace}

\journal{New Astronomy}

\def\astrobj#1{#1}
\begin{document}
\begin{frontmatter}

\title{Elemental abundances of the supergiant stars $\sigma$ Cygnus and $\eta$ Leonis}
\author[NU]{T. Tanr{\i}verdi\corref{cor2}}

\cortext[cor2]{e-mail: ttanriverdi@nigde.edu.tr}
\cortext[cor2]{This study was completed in Ankara University.}

\address[NU]{Ni\u{g}de University, Faculty of Arts and Sciences, Department of Physics, TR-51240, Ni\u{g}de, Turkey}

\begin{abstract}
\label{abstracts}

This study aims to analyse the elemental abundances for the late B 
type supergiant star \astrobj{$\sigma$ Cyg} and the early A-type supergiant  
\astrobj{$\eta$ Leo} using ATLAS9 \citep{Kurucz1995,sbordone2004}, assuming local thermodynamic equilibrium (LTE). The spectra used in this study are obtained from Dominion Astrophysical Observatory and have high resolution and signal-to-noise ratios. The effective temperature and the surface 
gravity of \astrobj{$\sigma$ Cyg} are determined from the ionisation equilibria of Al I/II, Mg I/II, Fe I/II, Fe II/III , and by fitting to the wings of H$_\gamma$ and H$_\beta$  profiles as \textit{T}$_{eff}$ = 10388 K and log \textit{g} = 1.80. The elemental abundances of \astrobj{$\eta$ Leo} are determined using \textit{T}$_{eff}$ = 9600 K and log \textit{g} = 2.00, as reported by \citet{P2006}.

The ionisation equilibria of C I/II, N I/II, Mg I/II, Ca I/II, Cr I/II and Fe I/II/III are also satisfied in the atmosphere of \astrobj{$\eta$ Leo}. The radial velocities of \astrobj{$\sigma$ Cyg} and \astrobj{$\eta$ Leo} are -7.25$\pm$7.57 km s$^{-1}$ and 10.40 $\pm$ 13.37 km s$^{-1}$, respectively. The derived projected rotational velocities \textit{vsini} from synthetic spectra are 27 and 2 km s$^{-1}$ for both stars, respectively. The macroturbulent velocities ($\zeta$) are 24 $\pm$ 2 km s$^{-1}$ and 14.5 $\pm$ 1.5 km s$^{-1}$. Also, the microturbulent velocities ($\xi$) have been determined for both of stars as 3.5 km s$^{-1}$. 
The CNO abundance results of \astrobj{$\sigma$ Cyg} and \astrobj{$\eta$ Leo} show C deficiency, N overabundance  and O in excess. 

\end{abstract}

\end{frontmatter}

\section{Introduction}
\label{introduction}
\sloppy

Late B and early A-type supergiants (hereafter-SGs) are visually luminous stars in our galaxy and other galaxies. Therefore, they are suitable candidates for further study. BA-SGs have been previously studied by many authors.

The comprehensive study of \citet{groth1961} was the first study reporting the temperature and chemical abundances of \astrobj{$\alpha$ Cyg}. \citet{aydin1972} presented the microturbulent velocities of some elements and gave atmospheric parameters of \astrobj{$\alpha$ Cyg}. Further studies of SGs located in the Milky Way and the Magellanic Clouds were conducted by \citet{Przy1968,Przy1971,Przy1972} and \citet{wolf1972,wolf1973}. \citet{wolf1971}, \citet{lambert1988} and \citet{lobel1992} studied the optical region of \astrobj{$\eta$\,Leo} using LTE methods and calculated the elemental abundances.

More recently, the chemical abundances of over twenty Galactic A-type SGs were calculated using progressed model atmospheres by \citet{venn1995a,venn1995b} and \citet{auf2002}. Subsequently, \citet{P2006} used very detailed nLTE line formation calculations to determine atmospheric parameters and elemental abundances (see also \citealt{P2000, Pet2001a, Pet2001b, Pb2001}). The chemical abundances of A-type SGs in many Local Group galaxies such as the, SMC (\citealt{venn1999}), M33 (\citealt{McCarthyetal1995}), M31 (\citealt{venn2000}), NGC 6822 (\citealt{venn2001}), WLM (\citealt{venn2003}) and \astrobj{Sextans A} (\citealt{kaufer2004}) also provide important clues about the chemical compositions of other galaxies. The quantitative spectroscopy of A-type SGs, beyond the Local Group Galaxies such as those located in NGC 3621 and NGC 300  were also presented\citep{bresolin2001,bresolin2002}.  Further, the possibility of the quantitative analysis of supergiants in Virgo cluster using the present generation of telescopes and new model atmospheres were firstly reported by \citet{Kud1995} and \citet{Kud1998}.

Spectral analysis of a prototype, A-type SG, \astrobj{$\alpha$ Cyg} was performed by \citet{alb2000} in LTE. \citet{auf2002} obtained the fundamental parameters and the mass loss rate of \astrobj{$\alpha$ Cyg} using PHOENIX, which computes line blanketed, nLTE atmospheric structures and synthetic spectra with winds. \citet{tt2004} determined preliminary abundance results of \astrobj{$\eta$ Leo} in LTE using ATLAS9. \citet{y2005} gave the elemental abundances of a late B and early A-type SGs (\astrobj{4 Lac} and \astrobj{$\nu$ Cep}). \citet{sp2008} presented nLTE elemental abundances of \astrobj{$\alpha$ Cyg} in detail in a recent study. \citet{mp2008} also investigated early and late B-type SGs. Recently, \citet{P2010} reported CNO abundances of more than ten BA-type SGs in our Galaxy. Additionaly, the spectral atlas of O9-A1.5-type SGs \citep{chentsov2007} and Deneb \citep{alb2003} has been published recently. 
\sloppy

The chemical analysis of late B and early A-type SGs are important in many respects, which are discussed below:

Their spectra are clear and exhibit a wide variety of chemical species including light elements (H, B, CNO), alpha elements (Mg, Si, S, Ca), iron-group elements (Sc, Ti, Cr, Mn, Fe, Ni) and s-process elements (Sr, Zr, Ba)\citep{P2002,P2006,venn1998}. Because the absorption lines of both $\alpha$ process and iron group elements are present in their spectra, A-type SGs are important for the determination of reliable [$\alpha$/Fe] ratios (see. \citealt{venn2003}).

Another reason is that they are the visually brightest stars in our galaxy and other galaxies. This characteristic makes them potential candidates for use in determining distances using their wind momentum - luminosity relation (hereafter WLR) \citep{Puls1996,Kud1999} and flux-weighted gravity-luminosity relation (hereafter FGLR) \citep{Kud2003, KudPrz2003}.

Notably, \citet{weiss2008} declared that "The results of nucleosynthesis in stars depend both on the conditions inside the star (temperature, density and chemical composition) and on the nuclear reaction rates. The accurate determination of elemental abundances therefore helps us to determine the interior stellar conditions, and properties of nuclei that are otherwise inaccessible." The elemental abundances of He, C, N, and O (C/N and N/O ratios) constitute an opportunity to test models of non-rotating (e.g. \citealt{schaller92}), as well as rotating and non-rotating models with mass loss (\citealt{ekstrom2012}).

The main goal of this study is to present the elemental abundances of \astrobj{$\sigma$ Cyg} and \astrobj{$\eta$ Leo} in LTE approximation. The suitability of, ATLAS9 model atmospheres for low luminosity SGs was shown by \cite{P2002} and mentioned by \citet{kaufer2004}.

\subsection{\astrobj{$\sigma$\,Cyg} and \astrobj{$\eta$\,Leo} }
\label{target stars}

\astrobj{$\sigma$\,Cyg} (HD202850, HIP 105102, SAO 71155) is a member of  the Cyg OB4 association \citep{hum78}. The galactic latitude and longitude of \astrobj{$\sigma$\,Cyg} are l = 84.1943 and b = -06.8723 respectively. It is classified as B9 Iab  \citep{morgan1953}. 

\astrobj{$\eta$\,Leo} (HD 87737, HIP 49583, SAO 98955) is an MK Standard and classified as A0 Ib \citep{morgan1950}. It is one of the brightest stars in the southern sky in the visual region of the spectrum. \astrobj{$\eta$\,Leo} is also a field star \citep{Bla1989} with the galactic coordinates l= 219.5301 b = +50.7501. The photometric variability of \astrobj{$\eta$\,Leo} was determined to be 0.${^m}$06 by \citet{adal2001}. The magnetic field of \astrobj{$\eta$\,Leo} was given by \citet{b2003} as 102.5 $\pm$ 59 Gauss. The UV spectrum of \astrobj{$\eta$\,Leo} was analysed by \citet{kon1976}, \citet{lamers1978} and \citet{Pra1980}. The mass loss rate was calculated by \citet{kon1976} from the resonance lines of Mg II lines and by \citet{bar1977} from the infrared excess as 3 $\times$ 10$^{-10}$  $M_\odot$ yr$^{-1}$ and 4.7 $\times$ 10$^{-8}$ $M_\odot$ yr$^{-1}$, respectively. 

The effective temperature ($T_{\rm eff}$), surface gravity (log\textit{g} \textit{in cgs.}), microturbulence ($\xi$) and macroturbulent velocity ($\zeta$) of \astrobj{$\sigma$\,Cyg} and \astrobj{$\eta$\,Leo} which have been previously determined by many authors are summarised in Table \ref{table1} with the methods they used.

\begin{table*}[ht]
\scriptsize
\caption{Stellar parameters of \astrobj{$\sigma$\,Cyg} and \astrobj{$\eta$\,Leo} from various sources.}
\label{table1}
\setlength{\tabcolsep}{.05cm}
\centering
\begin{tabular}{l c c c c c}
\hline\hline
Source						& $T_{\rm eff}$ in K	& $\log g$	& $\xi$	in km\,s$^{-1}$	& $\zeta$	in km\,s$^{-1}$ & Method\\
\hline
\multicolumn{4}{c}{{\underline{     $\sigma$ Cyg     }}}\\[3mm] 
\citet{P2010}&10800\,$\pm$\,200	& 1.85\,$\pm$0.10	& 6\,$\pm$\,1 & 35 &H, He lines,ionisation equilibria, SED\\[2mm]
\citet{mp2008}&   11000\,$\pm$\,500	& 1.85	& 7 & 33 & Balmer lines, SiII/SiIII\\[2mm]
\citet{OD1982}				& 10800		&--		& -- &--&--\\[2mm]
\citet{il1988}				& 11500\,$\pm$\,250			& 1.80\,$\pm$\,0.2	& 8.5\,$\pm$\,1 &--&H$\gamma$, H$\delta$ spectrophotometry, [c$_1$] index\\[2mm]
\multicolumn{4}{c}{{\underline{     $\eta$ Leo     }}}\\[3mm]
\citet{cenarro2007}& 9730\,$\pm$\,150	& 1.97	& 4\,$\pm$\,1 &--&--\\[2mm]
\citet{P2006}				& 9600\,$\pm$\,150	& 2.00\,$\pm$0.15	& 4\,$\pm$\,1 &16\,$\pm$\,2&Balmer lines,nLTE Mg I/II \\[2mm]
&\multicolumn{4}{c}{} & spectrophotometry\\[2mm]

\citet{venn1995a}				& 9700\,$\pm$\,200			& 2.0\,$\pm$\,0.2		& 4  &--& H$\gamma$\,, nLTE Mg I/II\\[2mm]
\citet{lobel1992}               & 10200\,$\pm$\,370			& 1.9\,$\pm$\,0.4		&5.4\,$\pm$\,0.7&--& LTE Fe I/II\\[2mm]
\citet{lambert1988}				& 10500			& 2.20		&3&--& H$\beta$\,, Str\"omgren photometry\\[2mm]
\citet{wolf1971}		& 10400\,$\pm$\,300		& 2.00\,$\pm$0.20& 2-10 &--& H$\beta$,H$\gamma$\,H$\delta$, Balmer jump, LTE Mg I/II,FeI/II\\[2mm]

\hline
\end{tabular}
\end{table*}




\section{The Spectra}
\label{thespectra}

The spectra of \astrobj{$\sigma$\,Cyg} (17 spectrograms) and \astrobj{$\eta$\,Leo} (17 spectrograms) were obtained at DAO (Dominion Astrophysical Observatory) by Dr. Saul J. Adelman. The wavelength coverages of the spectra are approximately $\lambda\lambda$\,3830-5210 and, $\lambda\lambda$\,5580-6740 for \astrobj{$\sigma$\,Cyg} and $\lambda\lambda$\,3800-6680 and, 6610-6740 for \astrobj{$\eta$\,Leo}. The reciprocal line dispersions of the spectra are 2.4 {\AA} mm$^{-1}$ for, $\lambda$ $<$ 6500 {\AA}  and 4.8 {\AA} mm$^{-1}$ for, $\lambda$ $>$ 6500 {\AA}, for the SITe-2 and SITe-4 detectors, respectively. Pixel-to-pixel resolution is 0.072 {\AA}, and the corresponding resolving power at 4500 {\AA} is R = 62500 . The signal-to-noise ratios (S/N) of the spectra are approximately 200-300. The spectra were rectified using the interactive graphical program REDUCE \citep{hf86} and line measurements were made using the graphical interface program VLINE \citep{hf86}. The scattered light was corrected with CCDSPEC \citep{gul1}.

The projected rotational velocities of \astrobj{$\sigma$\,Cyg} and \astrobj{$\eta$\,Leo} were determined from medium-strength lines to be 22.5 km s$^{-1}$ and 8.5 km s$^{-1}$, respectively. These values were used in VLINE as preliminary values for \textit{vsini} during measurements of the equivalent widths (EW) of spectral lines. 

\textit{A Multiplet Table of Astrophysical Interest} \citep{Moore1945} was the main source of line identifications. Other sources used included \citet{pet1983} for S II, \citet{hu192} for Ti II, \citet{ig1988} for V II, \citet{kiess1953} for Cr I, \citet{kiess1951} for Cr II, \citet{cat1964} for Mn I, \citet{ig1964} for Mn II, \citet{nave1994} for Fe I, \citet{dwo1971} and \citet{joh1978} for Fe II,  \citet{var1979} for Fe III, \citet{n1991} for Y II, and \citet{mcs1975} for singly ionised rare earth species. Previous studies of these types of stars were also used in the line identification (see \citealt{alb2003, gulliver2004, y2005, chentsov2007}). Both the identified stellar lines and the calculated abundances in the spectra are given in Table~\ref{table4}.

After heliocentric corrections for target stars, the radial velocities (hereafter -RV) were derived by comparing the stellar and laboratory wavelengths. The mean RV of \astrobj{$\sigma$\,Cyg} was found to be -7.25$\pm$7.57 km s$^{-1}$ with amplitude of 24.28 km s$^{-1}$. However, the amplitude of the RV in the atmospheres of \astrobj{$\sigma$\,Cyg} is closer to the macroturbulent velocity (24 $\pm$ 2 km s$^{-1}$) as in the case of other BA SGs (\astrobj{Deneb}, \astrobj{4 Lac} and \astrobj{$\nu$\,Cep}). The radial velocity of \astrobj{$\eta$\,Leo} is 10.40 $\pm$ 13.37 km s$^{-1}$ with amplitude of 33.31 km s$^{-1}$. Previous studies have shown that the radial velocities of the SGs can be attributed to radial and non-radial pulsations \citep{kaufer1997} or to a close or unresolved companion (see \citealt{P2002}). However, this case is not valid for \astrobj{$\eta$\,Leo} due to the obtained RV amplitude. Values given in the literature for the radial velocity of η Leo include, 2.6$\pm$0.7 km s$^{-1}$ \citep{vanhoof1963}, 1.6$\pm$3.1 km s$^{-1}$ \citep{abt1970} and 1.4$\pm$0.4 km s$^{-1}$ \citep{gon2006}. Therefore, the result implies that there is no evidence for the binarity of \astrobj{$\eta$\,Leo} as mentioned by \citet{Blazit1977}.


\begin{figure*}
\psfig{file=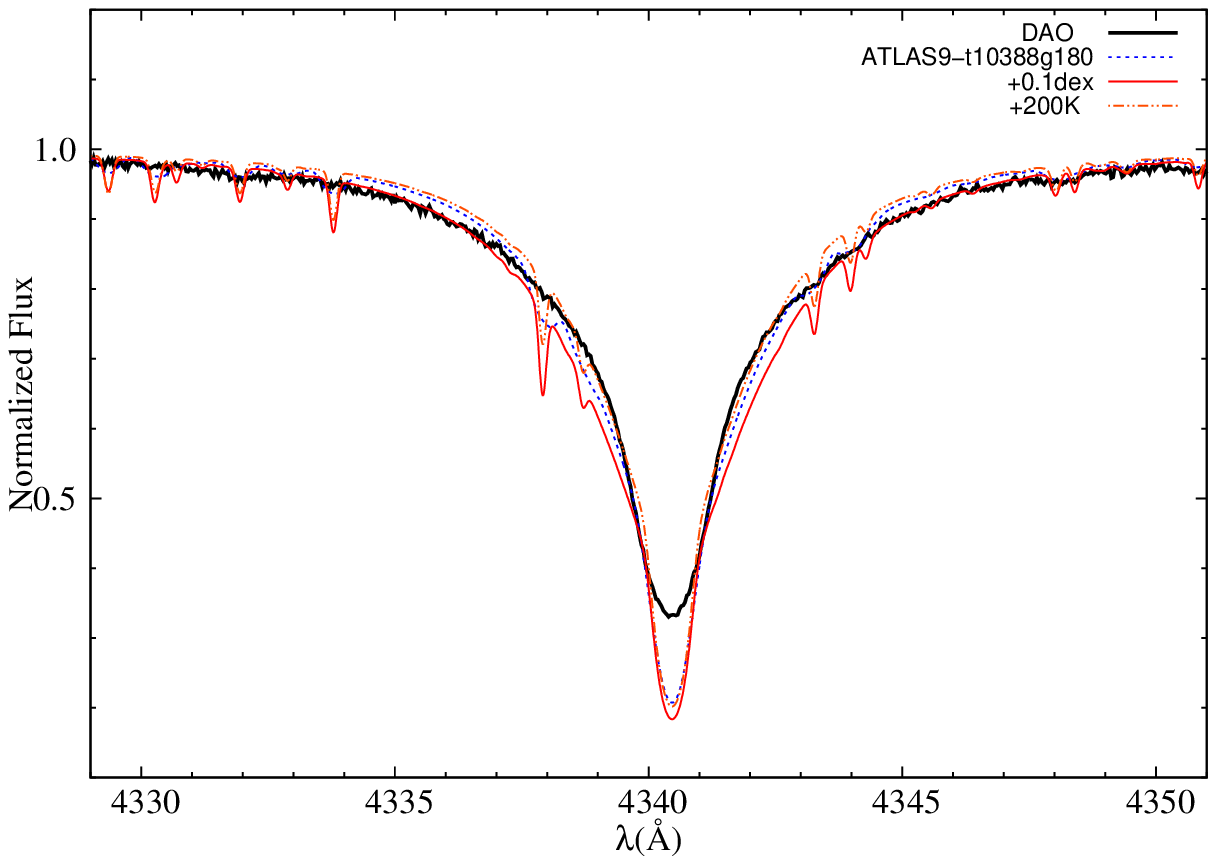,width=7.5cm}
\hskip 0.1cm
\psfig{file=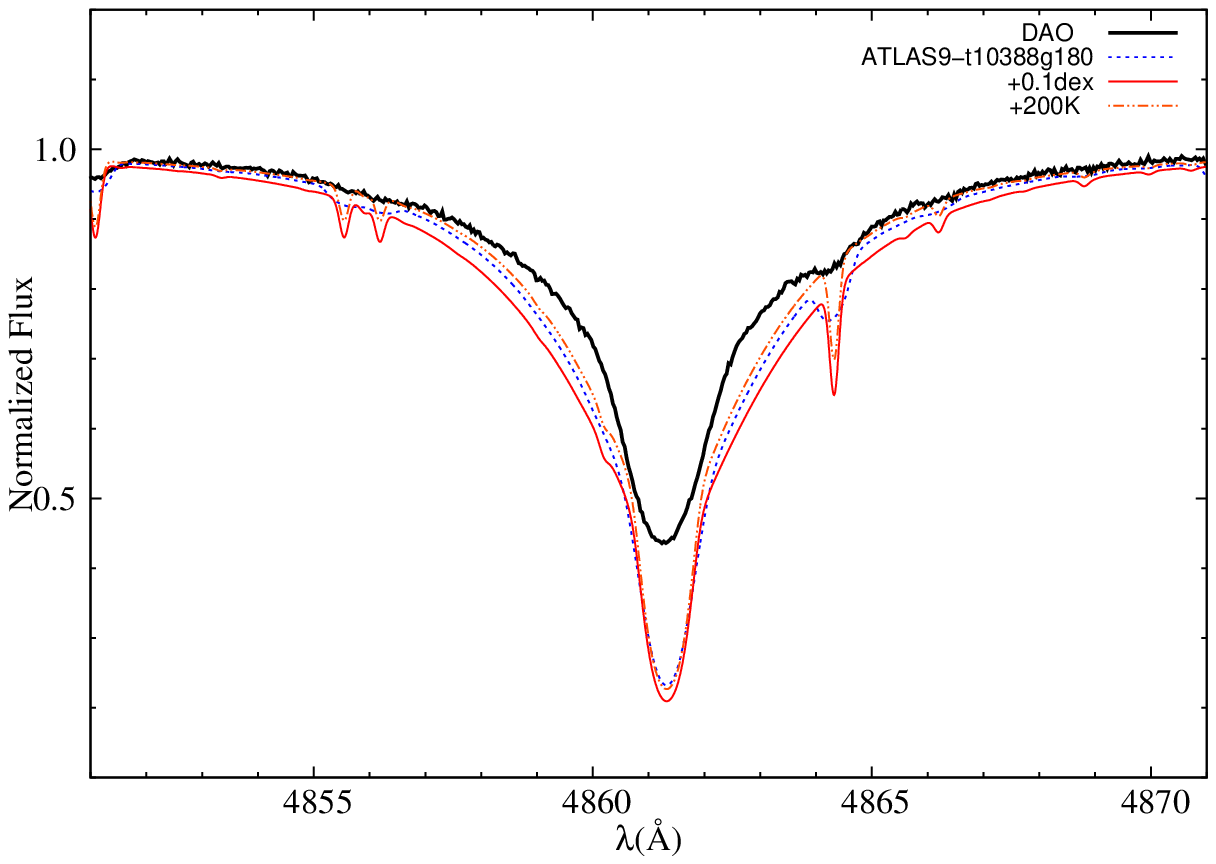,width=7.5cm}
\caption{The observed (black lines) H$_\gamma$ and H$_\beta$ profile with the synthetic fits (grey line) for \astrobj{$\sigma$ Cyg} at \textit{T}$_{eff}$ = 10388 K and log \textit{g} = 1.80.}
\label{fig1}
\end{figure*}

\begin{figure*}
\center
\includegraphics[width=120mm,height=70mm]{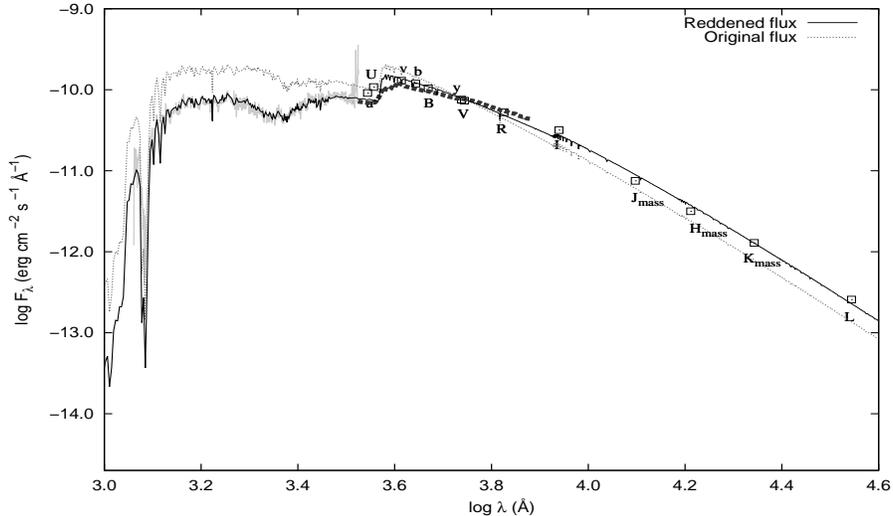}
\caption{A comparison of observed and computed fluxes (\textit{T}$_{eff}$ = 10388, log \textit{g} = 1.80) for \astrobj{$\sigma$ Cyg}; ATLAS9 model flux (dashed grey line), ATLAS9 reddened model flux (black line), IUE spectra (grey line), the spectrophotometric data of \citet{khar1998}(dotted line) and the photometric data (with squares) .}
\label{fig2}
\end{figure*}

\section{Stellar Parameters}
\label{stellarparameters}

The wings of Balmer lines are sensitive to both the effective temperature and gravity. The predicted synthetic profiles of H$_\gamma$ and H$_\beta$ are reproduced using SYNTHE \citep{Kurucz81} and matched with observations for several temperature-gravity pairs (Figure ~\ref{fig1},\ref{fig2}) until a consistent fit is found. In this study, ATLAS9 (\citealt{Kurucz1995,sbordone2004}) is used  assuming hydrostatic equilibrium, plane parallel geometry and LTE (local thermodynamic equilibrium) with solar metallicity [Fe/H = 0.0] and 4 km s$^{-1}$ microturbulent velocity. ATLAS9 models also include a very detailed line blanketing and important opacity sources \citep{Kurucz93}.

Another locus of temperature-gravity parameter pairs can be determined using ionisation equilibria in which equal abundances are derived from any consecutive ionic state. In this study, the ionisation equilibria are calculated for neutral and singly ionised species such as; Mg I/II, Al I/II and Fe I/II. In the case of \astrobj{$\sigma$ Cyg}, the abundance differences between the consecutive ionisation of Mg I/II, Al I/II, Fe I/II and Fe II/III are found to be of 0.11, 0.02, 0.12 and 0.01 dex, respectively.

\begin{figure}[h]
\centering
\includegraphics[width=90mm,height=60mm]{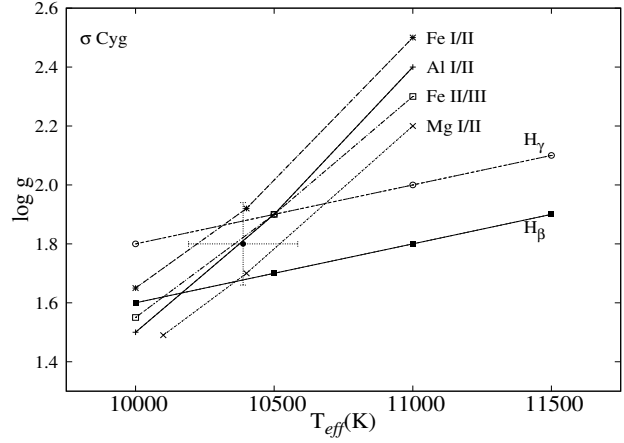}
\caption{The adopted value of \textit{T}$_{eff}$-log \textit{g} using Balmer line-wing and ionisation equilibria for \astrobj{$\sigma$ Cyg}}
\label{fig3}
\end{figure}

Theoretical fits to the wings of Balmer lines and LTE ionisation equilibrium loci give the effective temperature \textit{T}$_{eff}$ = 10388 $\pm$ 197 K and surface gravity log \textit{g} = 1.80 $\pm$ 0.14 dex for \astrobj{$\sigma$ Cyg} using the Kiel diagram (see Figure~\ref{fig3}). \textit{T}$_{eff}$ = 9600 $\pm$ 150 K and log \textit{g} = 2.00 $\pm$ 0.15 are used for \astrobj{$\eta$ Leo} \citep{P2006}. Moreover, the C I/II, N I/II, Mg I/II, Ca I/II, Cr I/II and Fe I/II/III ionisation equilibria are satisfied for \astrobj{$\eta$ Leo}, see Table~\ref{table2}.

The angular diameter of \astrobj{$\sigma$ Cyg} was given as 0.80, 0.26 and 0.53 mas (mili-arcsec) by \citet{hert1922}, \citet{morgan1953} and \cite{wess1969}, respectively. The adopted value of the angular diameter is 0.44 mas to generate the spectral energy distribution (SED). The observed SED of \astrobj{$\sigma$ Cyg} is derived by using low resolution IUE spectra (LWR 11614, SWP 15099), as well as the spectrophotometric data of \citet{khar1998}, and the photometric data of Johnson \citep{Reed2003,joh1966}, Str\"{o}mgren \citep{haumer1998} and 2MASS \citep{cutri2003} photometric data. The computed fluxes and the spectrophotometric data are reddened using the emprical approach of \citet{car1989} assuming a ratio of extinction to colour excess $R_V$\,=$A_V/E(B-V)$= 3.1 and E(B-V)=0.19 \citep{P2012}. 
The zero-points reported by \citet{heber2001} are used to transform the various magnitudes into fluxes. It is assumed that \textit{y = V} to transform \textit{b-y, c$_{1}$} and \textit{m$_{1}$} indexes to \textit{u, v, y} and \textit{b} magnitudes. The computed fluxes for \textit{T}$_{eff}$ = 10388 and log \textit{g} = 1.80 are consistent with the SED (see Figure ~\ref{fig2}).

\section{Abundance Analysis}
\label{abundanceanalysis}
\subsection{The results of the present study}
\label{resultofpresentstudy}

Helium abundances were calculated using SYNSPEC \citep{hubeny1994} and the metal abundances were derived using WIDTH9 \citep{Kurucz93}. The metal line damping constants were taken from \citet{Kurucz95}. However, the blended lines were neglected during the abundances analysis. The microturbulent velocity of \astrobj{$\sigma$ Cyg} was determined from only the Fe II lines and the microturbulent velocity of \astrobj{$\eta$ Leo} was ascertained using the Fe II lines and Cr II lines.

The microturbulent velocity was determined by finding the value at which the correlation between the derived abundances and the equivalent widths ($\xi_{1}$) was minimised and the minimum scatter about the mean abundance ($\xi_{2}$) was obtained \citep{Black1982}. The microturbulent velocity of \astrobj{$\sigma$ Cyg} was found to be approximately 3.5 km s$^{-1}$ and that of \astrobj{$\eta$ Leo} was calculated from Fe II and Cr II lines as 3.45 and 3.55 km s$^{-1}$, respectively. Therefore, it can be seen that the mean value of microturbulent velocity was approximately 3.5 km s$^{-1}$ for both stars (see Table~\ref{table2}). These derived microturbulent velocities were then used to calculate the elemental abundances.

The rotational velocities (\textit{vsini}) and macroturbulent ($\zeta$) velocities of \astrobj{$\sigma$ Cyg} were determined from intermediate lines between $\lambda\lambda$ 4500-4540 by finding the best fit between the theoretical and observed spectra using SYNTHE \citep{Kurucz81}, and are 27 $\pm$ 5 and  24 $\pm$ 2 km s$^{-1}$, respectively. Those of \astrobj{$\eta$ Leo} are 2 $\pm$ 2 and 14.5 $\pm$ 1.5 km s$^{-1}$, respectively.


\begin{table*}
\footnotesize
\centering
\caption{Microturbulence determinations}
\label{table2}
\begin{tabular*}{1.0\textwidth}{llllllll}
\hline
 Star        & Element & number    & $\xi_1$         & log (N/N$_T$)      & $\xi_2$         & log (N/N$_T$)  & Reference   \\  
             &         & of lines  & km s $^{-1}$    &                    &  km s$^{-1}$    &                &              \\
\hline
\astrobj{$\sigma$ Cyg} & Fe II   & 100      & 3.50            & -4.47$\pm$0.17     &  3.50           & -4.47$\pm$0.17 & KX+N4       \\       
             &         & adopted   & 3.50            &                    &                 &                &              \\
             &         &          &                 &                    &                 &                &              \\

$\eta$   Leo & Fe II   & 39       & 3.40            & -4.65$\pm$0.17     &  3.50           & -4.65$\pm$0.17 & KX+N4        \\       
	         & Cr II   & 150      & 3.50            & -6.49$\pm$0.21     &  3.60           & -6.49$\pm$0.21 & MF+KX+NL     \\
             &         & adopted  & 3.50            &                    &                 &                &              \\
\hline
\multicolumn{8}{l}{References of gf-values: MF = \citet{fuhr1988}, KX = \citet{Kurucz95}}\\ 
\multicolumn{8}{l}{N4 = \citet{fuhr2006}, NL = \citet{nil2006}.}\\

\end{tabular*}
\end{table*}


The helium abundance of \astrobj{$\sigma$ Cyg} and \astrobj{$\eta$ Leo} were determined by Dr. Adelman using the program SYNSPEC \citep{hubeny1994}. Table~\ref{table3} presents the He/H ratios which were determined to be (0.14 $\pm$ 0.01 for \astrobj{$\sigma$ Cyg} and 0.14 $\pm$ 0.02 for \astrobj{$\eta$ Leo}). To calculate the log \textit{N}/H from log \textit{N/N(total)} an offset of 0.06 was used for both \astrobj{$\sigma$ Cyg} and \astrobj{$\eta$ Leo}. 


\begin{table}
\footnotesize
\centering
\caption{He/H ratios of \astrobj{$\sigma$ Cyg} and \astrobj{$\eta$ Leo}.}
\label{table3}
\begin{tabular}{lll}
\hline
$\lambda$ ({\AA}) & \astrobj{$\sigma$ Cyg} & \astrobj{$\eta$ Leo} \\ 
\hline
4009 & ...  &  0.12 \\
4026 & 0.13 &   ... \\
4169 & ...  &  0.17 \\
4437 & ...  &  0.14 \\
4471 & 0.13 &  0.15 \\
4713 & 0.14 &  ...  \\
4922 & 0.13 &  0.11 \\
\hline
Average &0.14&0.14 \\
Std.Dev.& 0.01& 0.02 \\
\hline
\end{tabular}
\end{table}

\begin{figure}[h]
\centering
\includegraphics[width=120mm,height=80mm]{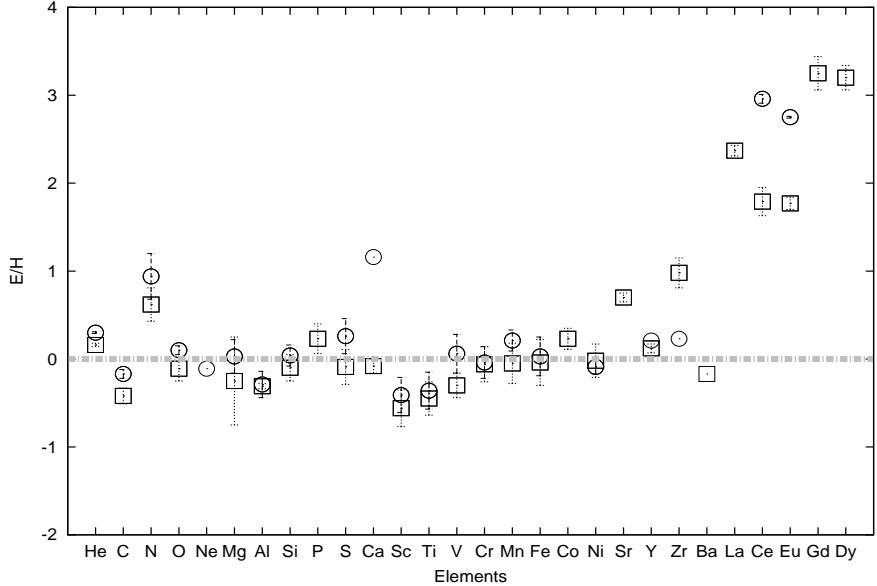}
\caption{Comparison with solar abundances. The open circles and squares indicate the relative  abundances of \astrobj{$\sigma$ Cyg} and \astrobj{$\eta$ Leo}, respectively.}
\label{fig4}
\end{figure}

Table~\ref{table4} includes the elemental abundances of the target stars, Table~\ref{table5} presents the error in our abundance analysis. For \astrobj{$\sigma$ Cyg}, carbon is deficient, nitrogen is overabundant and oxygen is near solar abundance. Neon is deficient by 0.12 dex, whereas magnesium and silicon are near solar abundance. \textit{Light elements}; aluminium and sulphur are underabundant, conversely, calcium is very overabundant. \textit{Iron group elements} are near solar abundance except for scandium, titanium and manganese. Scandium and titanium are underabundant, whereas, manganese is overabundant by 0.21 dex. Compared to solar values, the abundances of scandium, titanium, chromium, manganese, iron and nickel are -0.45, -0.36, -0.04, 0.19, 0.03 and -0.09 dex, respectively. \textit{Heavy elements}  tend to be over-abundant. Only, strontium is underabundant, whereas the abundances of yttrium and zirconium are comparable to their solar abundances. Caesium and europium are highly overabundant (see Figure~\ref{fig4}).

For \astrobj{$\eta$ Leo}, the CNO abundance show a pattern similar to that of \astrobj{$\sigma$ Cyg}. \textit{Light elements}, magnesium and aluminium are underabundant. The abundances of phosphorus matches the solar value, while the abundances of silicon, sulphur and calcium are near the solar value. \textit{Iron group elements} are also near solar abundance except for scandium, titanium and vanadium in \astrobj{$\eta$ Leo}'s atmosphere. Scandium and titanium and vanadium are underabundant whereas the abundances of chromium, manganese, iron, cobalt and nickel are solar. \textit{Heavy elements} tend to be overabundant. Although, strontium (-0.71 dex) and barium (-0.17 dex) are underabundant (see Figure~\ref{fig4}): alternatively, the abundances of these elements can be assumed to be solar according to \citet{ay2010}'s scale. 

\begin{table}[h] 
\footnotesize
\centering
\caption{Comparison of derived and solar abundances [N/H]} 
\label{table4}
\begin{tabular}{llllll}
\hline
Species & Sun$^{*}$ & \astrobj{$\sigma$ Cyg} &n  &\astrobj{$\eta$}   Leo & n    \\ 

\hline
He I  &10.99&   0.16$\pm$0.01  &4    &0.16$\pm$0.02    &5   \\ 
C I   & 8.55&   ...            &...  &-0.36$\pm$0.08   &3   \\ 
C II  & 8.55&   -0.19$\pm$0.05 &2    &-0.45$\pm$0.01   &1   \\  
N I   & 7.97&   ...            &...  &0.77$\pm$0.19    &8   \\ 
N II  & 7.97&   0.92$\pm$0.26  &8    &0.46$\pm$0.13    &4   \\  
O I   & 8.87&   0.08$\pm$0.05  &3    &-0.09$\pm$0.14   &4   \\
Ne I  & 8.08&   -0.13          &1    &...              &... \\   
Mg I  & 7.58&   -0.05$\pm$0.14 &3    &-0.31$\pm$0.07   &8   \\  
Mg II & 7.58&   0.06$\pm$0.13  &4    &-0.18$\pm$0.09   &11   \\  
Al I  & 6.47&   -0.29$\pm$0.11 &2    &-0.47$\pm$0.03   &2   \\  
Al II & 6.47&   -0.31$\pm$0.10 &4    &-0.15            &1   \\  
Si II & 7.55&   0.02$\pm$0.12  &5    &-0.06$\pm$0.18   &5   \\  
P II  & 5.45&   ...            &...  &+0.23$\pm$0.17   &4   \\  
S II  & 7.33&   0.24$\pm$0.20  &32   &-0.11$\pm$0.20   &17  \\  
Ca I  & 6.36&   1.35           &1    &-0.12            &1   \\
Ca II & 6.36&   0.93           &1    &-0.04$\pm$0.08   &3   \\  
Sc II & 3.17&   -0.39$\pm$0.20 &3    &-0.56$\pm$0.21   &7   \\  
Ti II & 5.02&   -0.34$\pm$0.21 &17   &-0.45$\pm$0.19   &56  \\  
V II  & 4.00&   0.06$\pm$0.22  &5    &-0.30$\pm$0.14   &17  \\  
Cr I  & 5.67&   ...            &...  &0.03$\pm$0.14    &6   \\  
Cr II & 5.67&   -0.06$\pm$0.18 &21   &-0.12$\pm$0.20   &35  \\  
Mn II & 5.39&   0.19$\pm$0.12  &10   &-0.06$\pm$0.20   &15  \\  
Fe I  & 7.50&   0.09$\pm$0.08  &7    & 0.00$\pm$0.20   &78  \\  
Fe II & 7.50&   -0.03$\pm$0.20 &98   &-0.11$\pm$0.17   &140 \\  
Fe III& 7.50&   -0.04          &1    &-0.01$\pm$0.14   &4   \\  
Co II & 4.92&    ...           &...  &0.23$\pm$0.12    &3   \\  
Ni II & 6.25&   -0.11$\pm$0.08 &3    &-0.02$\pm$0.19   &5   \\  
Sr II & 2.97&   -0.17$\pm$0.06 &2    &-0.71$\pm$0.05   &2   \\ 
Y II  & 2.24&   0.19		   &1    &0.12$\pm$0.15    &4   \\  
Zr II & 2.60&   0.21		   &1    &0.98$\pm$0.17    &7   \\  
Ba II & 2.13&   ...            &...  &-0.17            &1   \\
La II & 1.17&   ...            &...  &2.37$\pm$0.06    &2   \\  
Ce II & 1.58&   2.94$\pm$0.05  &3    &1.79$\pm$0.16    &3   \\  
Eu II & 0.51&   2.73$\pm$0.01  &2    &1.77$\pm$0.07    &2   \\                                                                  
Gd II & 1.12&   ...            &...  &3.19$\pm$0.14    &5   \\                                                                  
Dy II & 1.14&   ...            &...  &3.20$\pm$0.14    &3   \\                                                                  
\hline
\multicolumn{5}{l}{* \citet{Gre1996}}\\
\hline
\end{tabular}
\end{table} 

\begin{sidewaystable}[h] 
\small
\centering
\caption{Error sources for the abundances of the chemical elements of $\sigma$ Cyg.} 
\label{table5}
\begin{tabular}{lllllllll}
\hline
Ion&Abundances&$\sigma_{abn}$(scatter)& $\sigma_{abn}$(\textit{T}$_{eff}$)&$\sigma_{abn}$(log \textit{g})& $\sigma_{abn}$(v$_{turb}$)&$\sigma_{abn}$(\textit{gf-values})&$\sigma_{abn}$(EW)&$\sigma_{abn}$(syst.)\\ 
 & log (N/N$_T$) &                                  (dex) &             (dex)         &              (dex)         &             (dex)         &             (dex)&             (dex)&             (dex)             \\ 
 & ...              & ...                                  & (+200K)                     &    (+0.1dex)          &    ( +1 km s$^{-1}$)         &            (+$\%$10) &    ( +$\%$10 )       &  ...                        \\
\hline

C II	&-3.70	&0.05	&-0.09   &-0.04   &-0.06& -0.04  &+0.09  &0.15     \\  
N II    &-3.17	&0.26	&-0.12   &+0.05   &-0.04& -0.05  &+0.07  &0.16     \\
O I     &-3.11	&0.05	&-0.04   &-0.04   &-0.04& -0.05  &+0.06  &0.10     \\ 
Ne I    &-4.11	&...	&-0.07   &+0.05   &-0.05& -0.03  &+0.08  & 0.13     \\   
Mg I    &-4.53	&0.14	&+0.21   &-0.13   &-0.02& -0.02  &+0.05  &0.25     \\  
Mg II   &-4.42	&0.04	&+0.10   &-0.04   &-0.03& -0.04  &+0.06  &0.13     \\  
Al I    &-5.88	&0.11	&+0.12   &-0.07   &-0.01& -0.05  &+0.04  &0.15     \\   
Al II   &-5.90	&0.10	&-0.02   &+0.01   &-0.03& -0.04  &+0.05  &0.07     \\ 
Si II   &-4.49	&0.12	&+0.06   &+0.04   &0.00& -0.01  &+0.11  &0.14     \\  
S II    &-4.49	&0.20	&-0.06   &+0.04   &-0.05& -0.06  &+0.07  &0.15     \\  
Ca I    &-4.35	&...	&+0.27   &-0.11   & 0.00& -0.02  &+0.04  &0.29     \\ 
Ca II   &-4.77	&...	&+0.14   &-0.07   &-0.06& -0.01  &+0.09  &0.19     \\ 
Sc II   &-9.32	&0.20	&-0.09   &-0.04   &-0.00& -0.04  &+0.04  &0.11    \\ 
Ti II   &-7.42	&0.21	&+0.14   &-0.05   &-0.01& -0.09  &+0.06  &0.18     \\ 
V II    &-8.02	&0.22	&-0.09   &-0.04   &-0.01& -0.11  &+0.04  &0.15     \\ 
Cr II   &-6.45	&0.18	&+0.07   &-0.06   &-0.03& -0.20  &+0.16  &0.27     \\ 
Mn II   &-6.49	&0.12	&+0.07   &-0.03   &-0.00& -0.15  &+0.04  &0.17     \\  
Fe I    &-4.47	&0.08	&+0.23   &-0.05   &-0.01& -0.03  &+0.05  &0.24     \\ 
Fe II   &-4.59	&0.21	&+0.04   &-0.01   &-0.06& -0.18  &+0.06  &0.20     \\ 
Fe III  &-4.60	&...	&-0.06   &+0.06   &-0.05& -0.12  &+0.08  &0.17     \\ 
Ni II   &-5.92	&0.08	&+0.03   &+0.01   &-0.01& -0.27  &+0.05  &0.15    \\ 
Sr II   &-9.26	&0.06	&+0.14   &-0.06   &-0.01& -0.01  &+0.05  &0.16     \\
Y II    &-9.63	&...	&+0.10   &-0.11   &-0.00& -0.02  &+0.04  &0.16     \\ 
Zr II   &-9.25	&...	& 0.15   &-0.05   & 0.00&  0.00  &+0.05  &0.17     \\ 
Ce II   &-7.54	&0.05	&+0.13   &-0.06   &-0.06& -0.01  &+0.05  &0.16     \\ 
Eu II   &-8.84	&0.01	&+0.14   &-0.06   & 0.00& -0.02  &+0.05  &0.16     \\ 
\hline
\multicolumn{5}{l}{* \citet{Gre1996}}\\
\hline
\end{tabular}
\end{sidewaystable}

\newpage
\subsection{Comparison with previous studies}
\label{comparisonwithpreviousstudies}

\textit{\astrobj{$\sigma$ Cyg}}: The elemental abundance results of $\sigma$ Cyg are scarce in the literature. The first extensive abundance analysis of $\sigma$ Cyg was the pioneering study of \cite{il1988}. Although, \cite{il1988} used different temperature and surface gravity values, the abundance pattern determined in that study was similar to that of the present study, except that Fe is abundant and the abundance of Ti is solar. Helium is abundant in their study. It is given in Table~\ref{table6} with atmospheric parameters used and the number of lines. The measured EW, the \textit{gf} values used and the sources can be found in Table~\ref{table8}.

\citet{tt1995,tt1998, tt2000} also provide an analysis of the elemental abundances of \astrobj{$\sigma$ Cyg} partially in nLTE. Helium tends to be solar in their study, whereas Helium is abundant in the present study. They used the atmospheric parameters of \cite{il1988} in their study.

\citet{mp2008} also calculated the helium abundance as solar. In their study, Si was to be overabundant by 0.4 dex, as calculated using the Si II and Si III lines. Therefore, they identified $\sigma$ Cyg as a silicon star. Finally, \cite{P2010}'s reports helium is 0.38 by mass fraction and that CNO exhibits a similar to that presented in this paper.

The main advantages of the present analysis are the wide wavelength coverage, the high quality of our spectra and the signal-to-noise ratios. I also emphasise that not only the total number of ions (28) and elements (23), but also a greater number of lines per ion are analysed in the present study. Additionally, this is the first study reporting the heavy element abundances of $\sigma$ Cyg. However, nLTE effects are not considered in the present study.

\textit{\astrobj{$\eta$ Leo}}: The pioneering chemical abundance analysis of $\eta$ Leo was provided by \citet{wolf1971}. The results of \citet{wolf1971} indicated systematically larger abundance in this star due to its higher \textit{T}$_{eff}$.  Subsequently, \citet{venn1995a,venn1995b} reported elemental abundances in nLTE. Recently, $\eta$ Leo was studied extensively by \citet{P2002,P2006} (see Table~\ref{table7}). The most striking differences between their studies and this one is that, Sc and Ti are deficient. Because the analysis in this study was limited to nLTE calculations, however, the updated gf value of \citet{pickering2001,pickering2002} was used in the Ti abundance calculations. The heavy element abundances are similar to \citet{venn1995a}'s result. Furthermore, their elemental abundances of rare-earth and heavy elements (Sr, Zr, La, Ce, Eu, Gd and Dy) differ from the abundances given in the comprehensive studies of \citet{P2002,P2006} for $\eta$ Leo.  In the present study, we use the same \textit{T}$_{eff}$ and log \textit{g} with \citet{P2002} and determine the abundance of ions (35) and elements (27) using a greater number of lines.



\begin{sidewaystable}[h] 
\setlength{\tabcolsep}{.10cm}
\small
\centering
\caption{The comparison of derived \astrobj{$\sigma$ Cyg} abundances.} 
\label{table6}
\begin{tabular}{lllll}
\hline
Species   &This Study               &Ivanova \& Lyubimkov            &Takeda\& Takeda-Hidai          & \citet{mp2008}       \\                
          &                         &    
 (1988) 
         &(1995,1998 \& 2000)          &        \\ 
\hline                                                     
He I      &11.15\,$\pm$\,0.01(4)         &11.35\,$\pm$\,0.11(7)     &10.98(1)      &10.95\,$\pm$\,0.02()    \\   
C II      &8.36\,$\pm$\,0.05(2)          &8.25\,$\pm$\,0.10(3)      &8.13\,$\pm$\,0.09(2)    &...      \\  
N I       &...              &...          &8.46\,$\pm$\,0.01(7)    &...      \\  
N II      &8.89\,$\pm$\,0.26(8)          &8.97\,$\pm$\,0.08(4)      &...        &...      \\  
O I       &8.95\,$\pm$\,0.05(3)          &8.99(1)      &8.70(1)    &...      \\  
Ne I      &7.95(1)          &...          &...        &...      \\ 
Mg I      &7.53\,$\pm$\,0.14(3)	        &...          &...        &...      \\  
Mg II     &7.64\,$\pm$\,0.13(4)	        &7.80\,$\pm$\,0.06(4)      &...        &...      \\  
Al I      &6.18\,$\pm$\,0.11(2)	        &...          &...        &...      \\  
Al II     &6.16\,$\pm$\,0.10(4)	        &...          &...        &...      \\  
Si II     &7.57\,$\pm$\,0.12(5)	        &7.71\,$\pm$\,0.07(8)      &...        &         \\  
S II      &7.57\,$\pm$\,0.20(32)	        &7.50\,$\pm$\,0.06(11)     &...        &...      \\    
Ca I      &7.71\,$\pm$(1)          &...          &...        &...      \\   
Ca II     &7.29\,$\pm$(1)	        &...          &...        &...      \\  
Sc II     &2.72\,$\pm$\,0.20(3)	        &...          & ...       &...      \\  
Ti II     &4.73\,$\pm$\,0.21(17)	        &4.95\,$\pm$\,0.18(9)      &...        &...      \\    
V II      &4.06\,$\pm$\,0.22(5)	        &...          &...        &...      \\  
Cr II     &5.61\,$\pm$\,0.28(21)	        &5.70\,$\pm$\,0.07(10)     &5.91(2)    &...      \\    
Mn II     &5.58\,$\pm$\,0.12(10)	        &...          &...        &...      \\    
Fe I      &7.59\,$\pm$\,0.08(7)	        &...          &...        &...      \\  
Fe II     &7.48\,$\pm$\,0.20(98)         &7.73\,$\pm$\,0.03(77)     &7.51(4)    &...      \\  
Fe III    &7.46\,$\pm$(1)	        &...          &...        &...      \\  
Ni II     &6.14\,$\pm$\,0.08(3)	        &6.12\,$\pm$\,0.12(4)      &...        &...      \\  
Sr II     &2.80\,$\pm$\,0.06(3)	        &...      &...        &...      \\  
Y II      &2.43\,$\pm$(1)	        &...          &...        &...      \\   
Zr II     &2.81\,$\pm$(1)          &...          &...        &...      \\   
Ce II     &4.52\,$\pm$\,0.05(3)          &...          &...        &...      \\  
Eu II     &3.24\,$\pm$\,0.01(2)          &...          &...        &...      \\  
\hline
T$_{eff}$ (K)   &10388$\pm$\,197         &11500\,$\pm$\,200     & 11500\,$\pm$\,200 & 11000\,$\pm$\,1000   \\ 
log \textit{g} (cgs)&1.80$\pm$\,0.14      &1.80\,$\pm$\,0.20      & 1.80$\pm$\,0.20  & 1.85  \\                                                                       
\hline

\multicolumn{5}{l}{1. This Study, 2. Ivanova \& Lyubimkov (1988)}\\
\multicolumn{5}{l}{3. Takeda\& Takeda-Hidai(1995,1998 \& 2000) }\\
\multicolumn{5}{l}{4. \citet{mp2008},}\\
\end{tabular}
\end{sidewaystable}

\begin{sidewaystable}[h] 
\setlength{\tabcolsep}{.10cm}                                                                 
\small                                                             
\centering                                                                     
\caption{Comparison of derived \astrobj{$\eta$ Leo} abundances with previous studies.}           
\label{table7}                                                                 
\begin{tabular}{lllllll}                                                        
\hline                                                                         

Species    &This Study             &\citet{wolf1971}             &Venn         &     Takeda  &    Przybilla            \\         
           &                       &                             &(1995a,1995b)                          &\&Takeda-Hidai    & et al. (2006)\\  
           &                       &                             &                          &(1995,1998,2000)    & \\         
\hline                                                                
He I       &11.15\,$\pm$\,0.01(3)           &11.09(1)        &...        &10.70(1)       &11.18\,$\pm$\,0.04(14)          \\  
C I        &8.19\,$\pm$\,0.08(3)         &8.65(2)      &8.34\,$\pm$\,0.07(4)    &7.82\,$\pm$\,0.09(1)     &7.94\,$\pm$\,0.10(4)         \\  
C II       &8.10\,$\pm$\,0.01(1)         &...          &...        &8.13\,$\pm$\,0.10(4)        &8.10\,$\pm$\,0.09(3)	        \\  
N I        &8.74\,$\pm$\,0.19(8)         &...          &9.01\,$\pm$\,0.10(3)       &8.27(7)        &8.41\,$\pm$\,0.09(20)       \\ 
N II       &8.43\,$\pm$\,0.13(4)         &...          &...        &...         &8.32(1)         \\ 
O I        &8.78\,$\pm$\,0.14(4)         &8.83\,$\pm$\,0.67(7)      &8.97\,$\pm$\,0.10(6)    &8.70(3)        &8.78\,$\pm$\,0.09(9)         \\ 
Mg I       &7.27\,$\pm$\,0.07(8)         &7.51\,$\pm$\,0.22(3)      &7.55\,$\pm$\,0.05(4)    &...         &7.52\,$\pm$\,0.08(7)         \\ 
Mg II      &7.40\,$\pm$\,0.09(11)        &7.87\,$\pm$\,0.13(8)      &7.54\,$\pm$\,0.14(5)    &...         &7.53\,$\pm$\,0.04(12)        \\ 
Al I       &6.00\,$\pm$\,0.03(2)	        &6.45\,$\pm$\,0.13(2)          &...        &...         &6.11\,$\pm$\,0.06(2)         \\ 
Al II      &6.32(1)         &6.44(2)      &...        &...         &6.39\,$\pm$\,0.17(5)   	    \\ 
Si II      &7.45\,$\pm$\,0.18(5)         &8.03\,$\pm$\,0.14(9)      &...        &...         &7.58\,$\pm$\,0.19(4)   	    \\ 
Si III     &...             &...          &...        &...         &...    	        \\ 
S II       &7.24\,$\pm$\,0.20(17)        &...          &...        &...         &7.15\,$\pm$\,0.07(14)   	    \\ 
Ca I       &6.24(1)         &...          &...        &...         &...		        \\ 
Ca II      &6.32\,$\pm$\,0.08(3)         &6.16\,$\pm$\,0.11(2)      &...        &...         &6.31(1)   	    \\ 
Sc II      &2.61\,$\pm$\,0.21(6)         &2.97\,$\pm$\,0.24(5)      &...        &...         &2.57\,$\pm$\,0.14(3)   	    \\ 
Ti II      &4.57\,$\pm$\,0.19(56)        &4.73\,$\pm$\,0.34(82)          &4.77\,$\pm$\,0.21(21)   &...         &4.89\,$\pm$\,0.13(29)        \\ 
V II       &3.70\,$\pm$\,0.14(17)        &3.97\,$\pm$\,0.17(13)         &...        &...         &3.57\,$\pm$\,0.06(6)         \\ 
Cr II      &5.55\,$\pm$\,0.20(35)        &5.57\,$\pm$\,0.50(63)         &5.80\,$\pm$\,0.28(2)    &...         &5.62\,$\pm$\,0.08(29)        \\ 
Mn II      &5.33\,$\pm$\,0.20(15)        &4.66\,$\pm$\,0.12(6)         &5.40(1)    &...         &5.38\,$\pm$\,0.02(27)        \\ 
Fe I       &7.48\,$\pm$\,0.20(78)        &7.79\,$\pm$\,0.36(47)         &7.22\,$\pm$\,0.12(3)    &...         &7.34\,$\pm$\,0.12(21)        \\ 
Fe II      &7.40\,$\pm$\,0.17(140)       &7.76\,$\pm$\,0.25(73)         &7.47\,$\pm$\,0.18(19)   &...         &7.52\,$\pm$\,0.09(35)        \\ 
Fe III     &7.49\,$\pm$\,0.14(4)         &...          &...        &...         &...             \\ 
Co II      &5.11\,$\pm$\,0.12(3)         &3.65(1)         &...        &...         &...             \\ 
Ni II      &6.22\,$\pm$\,0.19(5)         &5.18\,$\pm$\,0.25         &...        &...         &6.30\,$\pm$\,0.06(7)         \\ 
Sr II      &2.26\,$\pm$\,0.05(2)         &...          &...        &...         &2.37\,$\pm$\,0.04(2)    	    \\ 
Y II       &3.48\,$\pm$\,0.15(4)         &...          &...        &...         &...  	        \\ 
Zr II      &3.44\,$\pm$\,0.17(7)         &...          &...        &...         &...  	        \\ 
Ba II      &1.96(1)         &3.63(1)         &...        &...         &2.00(1)  	    \\ 
La II      &3.54\,$\pm$\,0.06(2)         &...          &...        &...         &...   	        \\ 
Ce II      &3.37\,$\pm$\,0.16(2)         &...          &...        &...         &...   	        \\ 
Eu II      &2.28\,$\pm$\,0.07(2)         &...          &...        &...         &...  	        \\ 
Gd II      &4.31\,$\pm$\,0.14(5)         &...          & ...       &...         &...  	        \\ 
Dy II      &4.34\,$\pm$\,0.14(3)         &...          &...        &...         &...  	        \\ 
\hline
T$_{eff}$ (K)  &9600        &10400\,$\pm$\,300        &9700\,$\pm$\,200       &10200\,$\pm$\,300       &9600\,$\pm$\,150	        \\ 
log \textit{g} (cgs)&2.00   &2.05\,$\pm$\,0.2         &2.00\,$\pm$\,0.2       &1.90\,$\pm$\,0.2        &2.00\,$\pm$\,0.15	        \\                                                                       
\hline                                                                

\end{tabular}
\end{sidewaystable}


\section{Results and Discussion}
\label{Results}

\begin{figure*}[h]
\centering
\includegraphics[width=120mm,height=80mm]{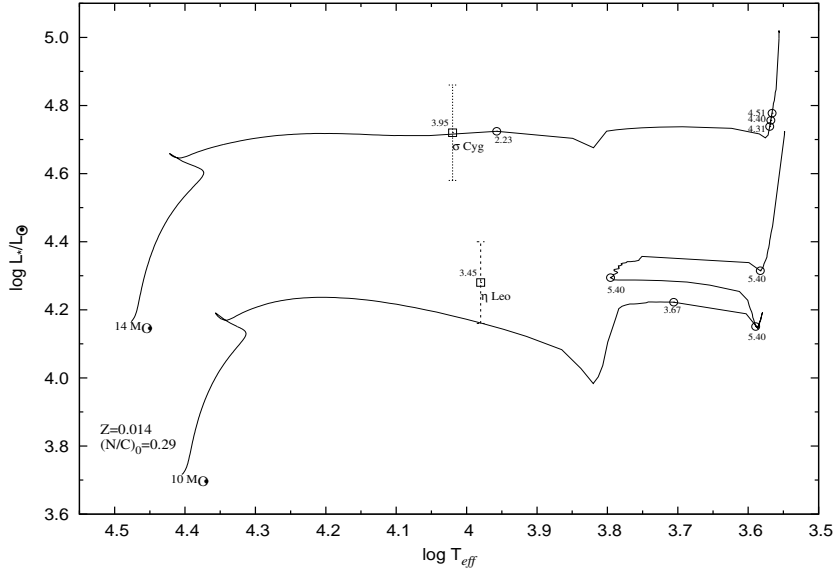}
\caption{Positions of $\sigma$ Cyg and $\eta$ Leo on HR Diagram for 10 $M_\odot$ and 14 $M_\odot$ using tracks of \citet{ekstrom2012}, the values of C/N ratios (by mass) give some points after MS (Main-Sequence) phase.} 
\label{fig5}
\end{figure*}

\renewcommand{\labelitemi}{$\bullet$}

Recent studies have shown that the microturbulent velocities of A-type SGs are in the range  4 - 8 km s$^{-1}$. For example, the derived microturbulent velocities of $\alpha$ Cyg (A2 Ia) are 7.5 km s$^{-1}$ (\citealt{alb2000}) and 8 $\pm$ 1 km s$^{-1}$ (\citealt{sp2008}), whereas the microturbulent velocity of $\nu$ Cep (A2 Ia) given in the literature is 5.2 km s$^{-1}$ (\citealt{y2005}). The microturbulent velocities of HD 111613 (A2 Iabe) and HD 92207 (A0 Iae) are 7 $\pm$ 1  and 8 $\pm$ 1 km s$^{-1}$, respectively (\citealt{P2006}). In this study, the determined microturbulent velocities are the same for \astrobj{$\sigma$ Cyg} and \astrobj{$\eta$ Leo}: 3.5 km s$^{-1}$. The obtained T$_{eff}$ and log \textit{g} values for \astrobj{$\sigma$ Cyg} are also consistent with those of \citet{P2010}. The T$_{eff}$ and log \textit{g} of \astrobj{$\eta$ Leo} are taken from the values determined by \citet{P2006}.

In the atmospheres of \astrobj{$\sigma$ Cyg}, The helium is abundant. C is depleted, N is strongly enriched and O is near solar abundance. The N/C and N/O abundances ratios (by mass) are 3.95 and 0.76 respectively, whereas the theoritical values of 4.31 and 0.77 for a rotating model of 14 $M_\odot$ calculated by \citet{ekstrom2012}. Theoritical values are given for initial value of blue loop(if any). The spectroscopic mass of \astrobj{$\sigma$ Cyg} is about 12 $M_\odot$. The initial value of \textbf{N/C} is 0.289. Light elements and iron group elements are near solar abundance. However, the abundances of heavy elements and rare-earth elements, except for Ba, are moderately overabundant.

For \astrobj{$\eta$ Leo}, C is deficient, N is in excess and O is near solar abundance. The abundance ratio of N/C and N/O (by mass) are 3.44 and 0.49, whereas the theoritical values are 5.40 and 0.82 for 10 $M_\odot$ by \citet{ekstrom2012}'s calculations. Light elements are also solar except for Mg and Al which are underabundant. Iron group elements are also near solar abundances except for Sc, Ti and V which are underabundant due to nLTE effects. However, the heavy elements are overabundant, except for Sr, which is underabundant and Ba, which is solar. 

The main contributors of s-process elements in the solar system (Sr, Y, Zr and Ba) are low to intermediate-mass asymptotic giant branch stars, whereas the r-process elements (Eu, Gd and Dy) are contributed by high mass Type II supernovae \citep{Travaglio2004}. The overabundances of some of the heavy elements are observed in the atmospheres of "normal" late B- and early-A- type stars. This overabundance can be referred to as the general enrichment of ISM by stellar and Galactic evolution and could be the topic of further investigation \citep{ad2004}.

\begin{figure*}[h]
\centering
\includegraphics[width=120mm,height=80mm]{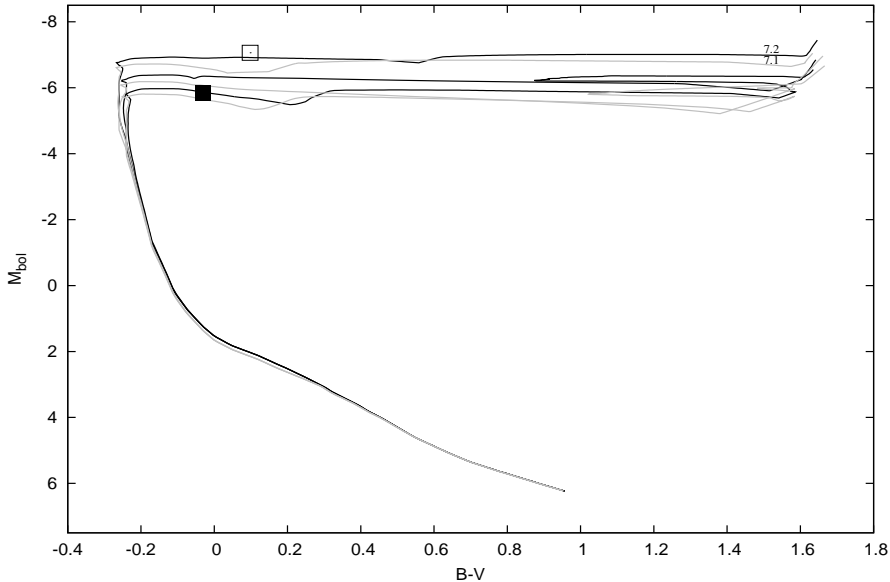}
\caption{ Isochrones of non-rotating (grey lines) and rotating models (black lines) of \citet{ekstrom2012}, $\sigma$ Cyg and $\eta$ Leo are represented by open and filled squares, respectively. Log ages are given by starting values of 7.1 and 7.2, respectively and increases with 0.1 increment for other lines.}
\label{fig6}
\end{figure*}

Theoretical evolutionary tracks, helium enrichment, CNO abundance patterns and light element abundances show that these stars have experienced the first dredge up phase as described by the tracks of \citet{schaller92} for 9$M_\odot$. This scenario is also cited by \cite{P2006} for \astrobj{$\eta$ Leo} referring to \cite{mm2003}. However, \cite{P2002} proposed that the result was caused by the accretion of nuclear processed matter from a redward-secondary component to the present faint primary and referred to the study of \cite{vanbeveren1998}. According to the stellar tracks of \cite{ekstrom2012}, a second scenario suggests that these stars evolve directly from the main sequence to red giant phase. This evolution was also proposed by \cite{venn1995a} referring to \cite{mm1989} based on helium enrichment. This scenario also supported binary nature as in the previous case(see Figure ~\ref{fig5}). \cite{ekstrom2012} described a 10$M_\odot$ scenario with rotation, which also supported blue loops. However, the possibility of binarity can not be ignored as noted by \cite{P2002}. The blue loops becomes shorten in the tracks of the masses larger than 9$M_\odot$ as it can be seen in Figure ~\ref{fig5}, therefore, it seems that \astrobj{$\sigma$ Cyg} have evolved directly from main-sequence to red-giant phase referring to \cite{ekstrom2012}, this scenario was also claimed by \citet{il1988}  and its progenitor should be an early-main sequence B star. And, \astrobj{$\eta$ Leo} has already experienced the first-dredge-up and should be on a blue-loop phase due to its helium enrichment and high N/C ratio.

The luminosity and temperature values used for \astrobj{$\sigma$ Cyg} are log \textit{L}/$L_\odot$ = 4.72 calculated using the FGLR \citep{Kud2003} and log \textit{T}$_{eff}$ = 4.02 (this study). Using log \textit{L}/$L_\odot$ = 4.28 and log \textit{T}$_{eff}$ = 3.98 \citep{P2006}. Figure~\ref{fig6} presents Log ages are near 7.2 and 7.4 for rotating models of these stars. Their ages correspond to 16 and 25 million years, respectively. The lower limit for the Cyg OB4 association age is 7 million years (\citealt{tetz2010}).  The N/C and N/O abundance ratios of \astrobj{$\sigma$ Cyg} and \astrobj{$\eta$ Leo} used are compatible with the values of Log age isochrones from the studies of \cite{ekstrom2012}.  \citet{il1988} found log \textit{L}/$L_\odot$ = 5.08 and log \textit{T}$_{eff}$ = 4.02, and estimated its age as 10 million years.

To determine the variability of the radial velocities in such stars more spectra spanning longer times are needed. Such observations allow us to determine the variability of these stars and the binary nature of \astrobj{$\eta$ Leo}. The elemental abundances of late B- and early A-type star indicate that it needs further refinement with new model atmospheres, updated atomic values such as new \textit{gf} values, and qualitative observations covering wide ranges of wavelengths with high resolution and S/N ratios. Such studies will allow us to obtain more accurate results for these types of stars and to determine the elemental abundances of heavy metals.


\section{Acknowledgements}
\label{acknowledgements}
I am thankful to James E. Hesser and the directory of the DAO (Dominion Astrophysical Observatory) for the observing time. This research utilised the SIMBAD database, which is operated at CDS, Strasbourg, France. I am also indebted to Dr. S. J. Adelman for contributing the spectra from DAO to help this study. I am grateful to Dr. B. Albayrak for his help during this study. Ankara University is also gratefully acknowledged for the support given to this research. A special thanks to the anonymous referee of this paper for careful reading of the manuscript and their generous comments.

\bibliographystyle{model2-names}

\appendix
\section{The analyses of metal lines}

\label{table8}

\begin{table*} 
\scriptsize                                                                     
\caption[ ]{Elemental Abundances of \astrobj{$\sigma$ Cyg} and \astrobj{$\eta$ Leo}}                      
\label{table8}
\begin{flushleft}                                                                   
\begin{tabular}{cccccrcrc}                                                      
    \hline                                                                          
 & & & & &\multicolumn{2}{c}{\astrobj{$\sigma$ Cyg}}&\multicolumn{2}{c}{\astrobj{$\eta$ Leo}}\\         
Species&Multiplet& $\lambda$(\AA)& log gf&Ref.&W$_{\lambda}$(m{\AA})                
&log N/N$_{T}$ &  W$_{\lambda}$(m{\AA})& log N/N$_{T}$ \\                           
 \hline                                                                             
                                                                                    
\noalign{\smallskip}                                                                
                                                                                    
C  I & & & & \multicolumn{3}{c}{log C/N$_{T}$ =} & &-3.87$\pm$0.08\\               
                                                  
& 6  & 4771.74 & -1.87 & FW & ...  &   ...  &   2.9& -3.85 \\                       
& 13 & 4932.05 & -1.66 & FW & ...  &   ...  &   3.8& -3.97 \\                       
& 14 & 4371.27 & -1.96 & FW & ...  &   ...  &   2.2& -3.79 \\                       
                                                                                    
\noalign{\smallskip}                                                                
                                                     
C II & & & & \multicolumn{3}{c}{log C/N$_{T}$ = -3.70$\pm$0.05} & &-3.96$\pm$0.03\\
                                                   
  &4  & 3918.97 & -0.53 & WF &  25.6& -3.76  & 12.0 & -3.97 \\
\end{tabular}                                                                                                                                                           
\end{flushleft}                                                                                                                                                         
\end{table*}                                                                                                                                                           
                                                                                    
\setcounter{table}{7}                                                               
\begin{table*}                                                                                                                                                      
\scriptsize  
\caption[ ]{- {\it continued}}                                                                                                             
\begin{flushleft}                                                                                                                                                      
\begin{tabular}{cccccrcrc}                                                          
                                                                                 
\hline                                                                              
                                                                                    
 & & & & &\multicolumn{2}{c}{\astrobj{$\sigma$ Cyg}}&\multicolumn{2}{c}{\astrobj{$\eta$ Leo}}\\         
Species&Multiplet& $\lambda$(\AA)& log gf&Ref.&W$_{\lambda}$(m{\AA})                
&log N/N$_{T}$ &  W$_{\lambda}$(m{\AA})& log N/N$_{T}$ \\                           
                                                                                    
 \hline                                                                             
Mg II&&&&\multicolumn{3}{r}{log Mg/N$_{T}$ =-4.42$\pm$0.13}&&-4.66$\pm$0.17\\                                                            
                                                                                    
  & 5  & 3848.24 & -1.57 & FW &  ... &...     &  47.5& -4.55\\                      
  &    & 3850.39 & -1.88 & FW &  ... &...     &  27.2& -4.57\\                      
  & 9  & 4428.00 & -1.21 & WS &  22.9& -4.52  &  21.2& -4.75\\                      
  &    & 4430.59 & -0.53 & KX &  70.0& -4.46  &  ... &   ...\\                      
  &    & 4433.99 & -0.91 & FW &  ... &...     &  41.3& -4.67\\
\end{tabular}                                                                                                                                                           
\end{flushleft}                                                                                                                                                         
\end{table*}                                                                        
                                                                                    
\setcounter{table}{7}                                                                                                                                                   
\begin{table*}                                                                                                                                                          
\scriptsize  
\caption[ ]{- {\it continued}}                                                                                                                                          
\begin{flushleft}                                                                                                                                                      
\begin{tabular}{cccccrcrc}                                                          

\hline                                                                              
                                                                                    
 & & & & &\multicolumn{2}{c}{\astrobj{$\sigma$ Cyg}}&\multicolumn{2}{c}{\astrobj{$\eta$ Leo}}\\         
Species&Multiplet& $\lambda$(\AA)& log gf&Ref.&W$_{\lambda}$(m{\AA})                
&log N/N$_{T}$ &  W$_{\lambda}$(m{\AA})& log N/N$_{T}$ \\                           
                                                                                    
 \hline                                                                             

S II &{\it (continued)}&    &       &       &      &       &      & \\                                                                               
\noalign{\smallskip}                                                                

 &17  & 6397.36 & -1.02 & KX &  15.8& -4.28  & ...  &  ...   \\                     
 &29  & 3993.50 & -0.82 & KX &  11.5& -4.63  & ...  &  ...   \\                     
 &35  & 4391.82 & -0.56 & WS &  ... & ...    & 2.5  &  -4.65 \\                     
 &40  & 4524.95 & +0.08 & WS &  23.0& -4.63  & 8.8  &  -4.96 \\                                                                                          
 &43  & 4456.39 & -0.56 & WS &   3.7& -4.74  & ...  &  ...   \\
\end{tabular}                                                                                                                                                           
\end{flushleft}                                                                                                                                                         
\end{table*}                                                                        
                                                                                    
\setcounter{table}{7}                                                                                                                                                   
\begin{table*}                                                                                                                                                         
\scriptsize  
\caption[ ]{- {\it continued}}                                                                                                                                          
\begin{flushleft}                                                                                                                                                      
\begin{tabular}{cccccrcrc}                                                          

\hline                                                                              
                                                                                    
 & & & & &\multicolumn{2}{c}{\astrobj{$\sigma$ Cyg}}&\multicolumn{2}{c}{\astrobj{$\eta$ Leo}}\\         
Species&Multiplet& $\lambda$(\AA)& log gf&Ref.&W$_{\lambda}$(m{\AA})                
&log N/N$_{T}$ &  W$_{\lambda}$(m{\AA})& log N/N$_{T}$ \\                           
                                                                                    
 \hline                                                                             
Ti II &{\it (continued)}&    &      &        &      &       &      & \\            
 &34  & 3913.46 & -0.42 & PT &  33.1& -7.71  &  86.6& -7.74 \\                      
 &40  & 4417.72 & -1.19 & PT &   8.2& -7.63  &  35.0& -7.61 \\                      
 &    & 4441.73 & -2.33 & PT & ...  &  ...   &   3.2& -7.60 \\                      
 &    & 4464.45 & -1.81 & PT & ...  &  ...   &  15.7& -7.48 \\                      
 &41  & 4300.06 & -0.44 & PT &  39.3& -7.58  & ...  &  ...  \\
\end{tabular}                                                                                                                                                          
\end{flushleft}                                                                                                                                                        
\end{table*}                                                                        
                                                                                    
\setcounter{table}{7}                                                                                                                                                   
\begin{table*}                                                                                                                                                          
\scriptsize  
\caption[ ]{- {\it continued}}                                                                                                                                          
\begin{flushleft}                                                                                                                                                      
\begin{tabular}{cccccrcrc}                                                          

\hline                                                                              
                                                                                    
 & & & & &\multicolumn{2}{c}{\astrobj{$\sigma$ Cyg}}&\multicolumn{2}{c}{\astrobj{$\eta$ Leo}}\\         
Species&Multiplet& $\lambda$(\AA)& log gf&Ref.&W$_{\lambda}$(m{\AA})                
&log N/N$_{T}$ &  W$_{\lambda}$(m{\AA})& log N/N$_{T}$ \\                           
                                                                                    
 \hline                                                                             
V II&&&&\multicolumn{3}{c}{log V/N$_{T}$ =-8.02$\pm$0.22}&&-8.36$\pm$0.14\\                  
                                                                                    
 & 9  & 3968.11 & -1.31 & BG & ...  &  ...   &   6.6& -8.39 \\                      
 &    & 3997.12 & -1.23 & BG &   5.2& -7.99  &   5.2& -8.53 \\                      
 &    & 4036.78 & -1.59 & BG & ...  &  ...   &   3.1& -8.41 \\                      
 &10  & 3916.42 & -1.05 & BG &   4.8& -8.24  &   9.4& -8.47 \\                      
 &11  & 3866.74 & -1.55 & BG & ...  &  ...   &   3.7& -8.38 \\
\end{tabular}                                                                                                                                                   
\end{flushleft}                                                                                                                                      
\end{table*}                                                                       
                                                                                    
\setcounter{table}{7}                                                                                                                                                   
\begin{table*}                                                                                                                                                         
\scriptsize  
\caption[ ]{- {\it continued}}                                                                                                                                          
\begin{flushleft}                                                                                                                                                       
\begin{tabular}{cccccrcrc}                                                          

\hline                                                                              
                                                                                    
 & & & & &\multicolumn{2}{c}{\astrobj{$\sigma$ Cyg}}&\multicolumn{2}{c}{\astrobj{$\eta$ Leo}}\\         
Species&Multiplet& $\lambda$(\AA)& log gf&Ref.&W$_{\lambda}$(m{\AA})                
&log N/N$_{T}$ &  W$_{\lambda}$(m{\AA})& log N/N$_{T}$ \\                           
                                                                                    
 \hline

Cr II &{\it (continued)}&    &      &        &      &       &      & \\ 
&129  & 3911.32 & -2.06 & KX &  ... &  ...   &   7.3& -6.36 \\                      
&130  & 3866.00 & -2.35 & KX &  ... &  ...   &   4.7& -6.28 \\                      
&     & 3866.52 & -2.07 & KX &  ... &  ...   &   9.4& -6.24 \\                      
&162  & 4145.77 & -1.16 & KX &  ... &  ...   &  27.2& -6.42 \\                      
&     & 4224.85 & -1.73 & KX &  10.8& -6.02  & ...  &   ... \\
\end{tabular}                                                                                                                                                          
\end{flushleft}                                                                                                                                                        
\end{table*}                                                                        
                                                                                    
\setcounter{table}{7}                                                                                                                                                   
\begin{table*}                                                                                                                                                        
\scriptsize  
\caption[ ]{- {\it continued}}                                                                                                                                         
\begin{flushleft}                                                                                                                                                      
\begin{tabular}{cccccrcrc}                                                          

\hline                                                                                                                                                                 
 & & & & &\multicolumn{2}{c}{\astrobj{$\sigma$ Cyg}}&\multicolumn{2}{c}{\astrobj{$\eta$ Leo}}\\         
Species&Multiplet& $\lambda$(\AA)& log gf&Ref.&W$_{\lambda}$(m{\AA})                
&log N/N$_{T}$ &  W$_{\lambda}$(m{\AA})& log N/N$_{T}$ \\                           
                                                                                    
 \hline                                                                             
Fe I &{\it (continued)}&    &       &       &      &       &      & \\
 & 43   & 4071.74 & -0.02 & N4 &  14.8& -4.40  & ...  &  ... \\                       
 &    & 4132.06 & -0.67 & N4 &  ... &  ...   &  11.6& -4.72\\                       
 & 68 & 4442.34 & -1.26 & N4 &  ... &  ...   &   1.9& -4.60\\                       
 &    & 4447.72 & -1.34 & N4 &  ... &  ...   &   3.1& -4.29\\                       
 &    & 4459.12 & -1.28 & N4 &  ... &  ...   &   3.5& -4.33\\
\end{tabular}                                                                                                                                                          
\end{flushleft}                                                                                                                                                         
\end{table*}                                                                        
                                                                                    
\setcounter{table}{7}                                                               
                                                                                    
\begin{table*}                                                                      
\scriptsize                                                                                 
\caption[ ]{- {\it continued}}                                                      
                                                                                    
\begin{flushleft}                                                                   
                                                                                    
\begin{tabular}{cccccrcrc}                                                          
\hline                                                                              
                                                                                    
 & & & & &\multicolumn{2}{c}{\astrobj{$\sigma$ Cyg}}&\multicolumn{2}{c}{\astrobj{$\eta$ Leo}}\\         
Species&Multiplet& $\lambda$(\AA)& log gf&Ref.&W$_{\lambda}$(m{\AA})                
&log N/N$_{T}$ &  W$_{\lambda}$(m{\AA})& log N/N$_{T}$ \\                           
                                                                                    
 \hline                                                                             
 Fe I &{\it (continued)}&    &       &       &      &       &      & \\ 
 
  &559 & 4067.98 & -0.47 & N4 &  ... &  ...   &   2.7& -4.62\\ 
 &561 & 4005.24 & -0.61 & N4 &  ... &  ...   &  10.4& -4.62\\
 &562 & 3948.10 & -0.56 & N4 &  ... &  ...   &   3.6& -4.38\\                                                                                                                               
 &661 & 3951.16 & -0.30 & N4 &  ... &  ...   &   1.8& -4.92\\                     
 &664 & 3846.80 & -0.02 & N4 &  ... &  ...   &   6.5& -4.64\\
\end{tabular}                                                                                                                                                         
\end{flushleft}                                                                                                                                                         
\end{table*}                                                                        
                                                                                    
\setcounter{table}{7}                                                               
                                                                                    
\begin{table*}                                                                      
\scriptsize                                                                                   
\caption[ ]{- {\it continued}}                                                      
                                                                                    
\begin{flushleft}                                                                   
                                                                                    
\begin{tabular}{cccccrcrc}                                                          
\hline                                                                              
                                                                                    
 & & & & &\multicolumn{2}{c}{\astrobj{$\sigma$ Cyg}}&\multicolumn{2}{c}{\astrobj{$\eta$ Leo}}\\         
Species&Multiplet& $\lambda$(\AA)& log gf&Ref.&W$_{\lambda}$(m{\AA})                
&log N/N$_{T}$ &  W$_{\lambda}$(m{\AA})& log N/N$_{T}$ \\                           
                                                                                    
 \hline                                                                             
Fe II &{\it (continued)}&    &      &       &      &       &      & \\            
 & 32 & 4439.13 & -5.27 & KX &  ... &  ...   &   2.1& -4.53\\ 
 & 33 & 4372.22 & -4.38 & KX &   3.1& -5.07  & ...  &  ... \\                       
 & 35 & 5161.18 & -4.48 & KX &  ... &  ...   &   9.4& -4.55\\            
 & 36 & 4993.35 & -3.68 & N4 &  ... &  ...   &  39.9& -4.65\\                                                                      
 & 37 & 4472.92 & -3.53 & N4 &  ... &  ...   &  41.5& -4.75\\
\end{tabular}                                                                                                                                                         
\end{flushleft}                                                                                                                                                       
\end{table*}                                                                        
                                                                                    
\setcounter{table}{7}                                                               
                                                                                    
\begin{table*}                                                                      
\scriptsize                                                                                      
\caption[ ]{- {\it continued}}                                                      
                                                                                    
\begin{flushleft}                                                                   
                                                                                    
\begin{tabular}{cccccrcrc}                                                          
\hline                                                                              
                                                                                    
 & & & & &\multicolumn{2}{c}{\astrobj{$\sigma$ Cyg}}&\multicolumn{2}{c}{\astrobj{$\eta$ Leo}}\\         
Species&Multiplet& $\lambda$(\AA)& log gf&Ref.&W$_{\lambda}$(m{\AA})                
&log N/N$_{T}$ &  W$_{\lambda}$(m{\AA})& log N/N$_{T}$ \\                           
                                                                                    
 \hline                                                                             
Fe II &{\it (continued)}&    &       &       &      &       &      & \\            
 &190 & 3996.35 & -2.93 & KX &   4.8& -4.57  &   3.2& -4.89\\  
 &212 & 3960.90 & -1.42 & KX &  21.3& -4.70  &  23.5& -4.76\\                                                                                                                                                                         
 &213 & 4354.34 & -1.74 & KX &  22.1& -4.16  &  19.1& -4.34\\                      
 &    & 4507.10 & -1.92 & KX &  11.4& -4.25  &   9.4& -4.43\\                                                                                                         
 &219 & 4598.53 & -1.50 & KX &  10.9& -4.67  & ...  &  ... \\
\end{tabular}                                                                                                                                                         
\end{flushleft}                                                                                                                                                       
\end{table*}                                                                        
                                                                                    
\setcounter{table}{7}                                                               
                                                                                    
\begin{table*}                                                                      
\scriptsize                                                                                      
\caption[ ]{- {\it continued}}                                                      
                                                                                    
\begin{flushleft}                                                                   
                                                                                    
\begin{tabular}{cccccrcrc}                                                          
\hline                                                                              
                                                                                    
 & & & & &\multicolumn{2}{c}{\astrobj{$\sigma$ Cyg}}&\multicolumn{2}{c}{\astrobj{$\eta$ Leo}}\\         
Species&Multiplet& $\lambda$(\AA)& log gf&Ref.&W$_{\lambda}$(m{\AA})                
&log N/N$_{T}$ &  W$_{\lambda}$(m{\AA})& log N/N$_{T}$ \\                           
                                                                                    
 \hline 
Fe II &{\it (continued)}&    &       &       &      &       &      & \\ 
 &    & 4974.22 & -0.84 & KX &   6.8& -4.26  & ...  &  ...  \\                      
 &    & 4977.03 &  0.04 & KX &  21.7& -4.55  & ...  &  ...  \\                      
 &    & 4977.92 & -0.67 & KX &   8.4& -4.33  & ...  &  ...  \\                      
                    
 &    & 4984.49 & +0.01 & KX &  25.9& -4.44  &  20.9& -4.58\\                      
 &    & 4990.50 & +0.18 & KX &  28.5& -4.55  &  28.8& -4.56 \\
\end{tabular}                                                                                                                                                         
\end{flushleft}                                                                                                                                                       
\end{table*}

\setcounter{table}{7}                                                               
                                                                                    
\begin{table*}                                                                      
\scriptsize                                                                                      
\caption[ ]{- {\it continued}}                                                      
                                                                                    
\begin{flushleft}                                                                   
                                                                                    
\begin{tabular}{cccccrcrc}                                                          
\hline                                                                              
                                                                                    
 & & & & &\multicolumn{2}{c}{\astrobj{$\sigma$ Cyg}}&\multicolumn{2}{c}{\astrobj{$\eta$ Leo}}\\         
Species&Multiplet& $\lambda$(\AA)& log gf&Ref.&W$_{\lambda}$(m{\AA})                
&log N/N$_{T}$ &  W$_{\lambda}$(m{\AA})& log N/N$_{T}$ \\                           
                                                                                    
 \hline 

\noalign{\smallskip}                                                                                                                                                   
Fe III&&&&\multicolumn{3}{c}{log Fe/N$_{T}$ = -4.60} & &-4.57$\pm$0.14\\            
                                                                                    
 &  4 & 4382.51 & -3.02 & KX & ...  &  ...   &   4.0& -4.44 \\                      
 &    & 4395.76 & -2.60 & KX & ...  &  ...   &   8.5& -4.44 \\                      
 &    & 4419.60 & -2.22 & KX &  21.5& -4.60  &   9.3& -4.76 \\                      
 &    & 4431.02 & -2.57 & KX & ...  &  ...   &   6.3& -4.63 \\                      
                                                                                    
\noalign{\smallskip}                                                                                                                                                  
Co II&&&&\multicolumn{3}{c}{log Co/N$_{T}$ = } & &-6.91$\pm$0.12\\
\end{tabular}                                                                                                                                                         
\end{flushleft}                                                                                                                                                       
\end{table*}

\setcounter{table}{7}                                                               
                                                                                    
\begin{table*}                                                                      
\scriptsize                                                                                      
\caption[ ]{- {\it continued}}                                                      
                                                                                    
\begin{flushleft}                                                                   
                                                                                    
\begin{tabular}{cccccrcrc}                                                          
\hline                                                                              
                                                                                    
 & & & & &\multicolumn{2}{c}{\astrobj{$\sigma$ Cyg}}&\multicolumn{2}{c}{\astrobj{$\eta$ Leo}}\\         
Species&Multiplet& $\lambda$(\AA)& log gf&Ref.&W$_{\lambda}$(m{\AA})                
&log N/N$_{T}$ &  W$_{\lambda}$(m{\AA})& log N/N$_{T}$ \\                           
                                                                                    
 \hline 
Gd II&&&&\multicolumn{3}{c}{log Gd/N$_{T}$ = ...} & &-7.75$\pm$0.14\\                
 
 &115 & 4463.24 & -0.97 & CB & ...  &  ...   &   8.3& -7.38\\ 
 &    & 4755.24 & -0.24 & CB & ...  &  ...   &   3.3& -7.93\\        
 &    & 5130.27 & -0.95 & CB & ...  &  ...   &   3.3& -7.66\\
 &    & 5191.80 & -0.94 & KX & ...  &  ...   &   3.0& -7.81\\
 &    & 5210.48 & -0.93 & KX & ...  &  ...   &   5.2& -7.58\\

Dy II&&&&\multicolumn{3}{c}{log Dy/N$_{T}$ = ...} & &-9.07$\pm$0.14\\                
 
 & -  & 3872.10 & +0.01 & WL & ...  &  ...   &   8.3& -9.00\\ 
 &    & 4077.96 & -0.06 & WL & ...  &  ...   &   3.5& -9.27\\        
 &    & 4103.30 & -0.38 & WL & ...  &  ...   &   3.5& -8.95\\
 &    &         &        &   &        &      &&\\   
\hline                                                                            
\noalign{\smallskip}                                                                                  
\end{tabular}\\

Note: gf value references follow: \\                                                
AT = \citet{aldenius2007};
B = \citet{brage1998};\\
BG = \citet{biemont1981} for Zr II, \citet{biemont1989} for V II;\\           
CB = \citet{corlis1962};
FW = \cite{fuhr2002} and \citet{fuhr1988};\\                                                                             
HL = \citet{hannaford1982};                                                                                                      
KG = \citet{kling2000};
KS = \citet{kling2001};\\
KX = \citet{Kurucz95};                                                     
LA = \citet{lanz1985};                                                               
LD = \citet{lawler1989};\\
LN = \citet{ljung2006};
LW = \citet{lawler2001};
NL = \citet{nil2006};\\
N4 = \citet{fuhr2006};
MF = \citet{fuhr1988} and \citet{martin1988};\\                                                                    
RP = \citet{raassen1998};
PT = \citet{pickering2001} and \citet{pickering2002};\\
SG = \citet{schulz1969};
WF = \cite{wiese1996};                     
WL = \cite{wickliffe2000};\\      
WM = \citet{wiese1980};                                                      
WS = \citet{wiese1969}; 
ZS = \citet{zhang2001};

\end{flushleft}                                                                     
\end{table*}   

\end{document}